# Random Access Transport Capacity

Jeffrey G. Andrews, Steven Weber, Marios Kountouris, Martin Haenggi

**Abstract**

We develop a new metric for quantifying end-to-end throughput in multihop wireless networks, which we term *random access transport capacity*, since the interference model presumes uncoordinated transmissions. The metric quantifies the average maximum rate of successful end-to-end transmissions, multiplied by the communication distance, and normalized by the network area. We show that a simple upper bound on this quantity is computable in closed-form in terms of key network parameters when the number of retransmissions is not restricted and the hops are assumed to be equally spaced on a line between the source and destination. We also derive the optimum number of hops and optimal per hop success probability and show that our result follows the well-known square root scaling law while providing exact expressions for the preconstants as well. Numerical results demonstrate that the upper bound is accurate for the purpose of determining the optimal hop count and success (or outage) probability.

## I. Introduction

Determining the capacity of distributed wireless networks (*i.e.,* ad hoc networks) is one of the most general and challenging open problems in information theory. Straightforward applications of known information theoretic tools and inequalities becomes intractable almost immediately and have hence yielded little in the way of results. This motivates the exploration of approaches to describing ad hoc network throughput that, while falling short of strict information theory upper bound standards, do provide insight into the fundamental trends on achievable throughput.

J. Andrews is with the University of Texas at Austin, S. Weber is with Drexel University, M. Kountouris is with the SUPELEC, and M. Haenggi is with Notre Dame University. The contact author is J. Andrews, jandrews@ece.utexas.edu. This research has been supported primarily by the DARPA Information Theory for Mobile Ad Hoc Networks (IT-MANET) program. It has appeared in part at the 2009 Allerton conference. Manuscript date: October 31, 2018.

### A. Motivation and Related Work

The main line of present inquiry is to consider the *transport capacity* of an ad hoc network, which quantifies the bits per second that can be reliably communicated over some distance in the network [1]. A significant merit of this approach is that it describes important aspects of the capacity region of all possible rate pairs in the network – notably how the region scales with the number of nodes $n$ – and has provided high-level insight on how different network scenarios and approaches may affect the scaling law. Notwithstanding special cases that suggest more optimistic scalings [2], [3], [4], it is generally agreed that the scaling law in ad hoc networks is $\Theta(\sqrt{n})$ and can be achieved with nearest-neighbor routing through the network [5], [6], [7], [8], [9]. Despite this considerable progress, a common limitation of this line of research is that the main results are typically asymptotic scaling laws that do not easily allow capacity tradeoffs to be made based on the salient network parameters and design choices. Although many efforts have been made to better understand the transport capacity preconstants, e.g. [10], [11], [12], in general tractability has suffered and/or several essential aspects of ad hoc networks have had to be abstracted out.

An alternative approach pioneered by the current authors and others has focused on computing achievable rate regions and network densities for different architectures and technologies. A key to this approach is to assume that the interferer locations are random, in particular that they are Poisson distributed over the plane. Although this is an idealization, it accurately models the scenario where transmitters are randomly scattered and uncoordinated, which seems to be quite a bit more realistic than other popular alternatives, such as assuming they are placed deterministically on a regular grid. This approach has allowed closed-form derivation of outage probability as well as the maximum allowable transmission density at a specified outage probability and data rate, the latter of which we have termed the *transmission capacity* [13], [14], [15]. The analytical tractability of this approach (using tools from stochastic geometry [16], [17]) has to date allowed quantitative design tradeoffs for spread spectrum [13], [18], [19], interference cancellation [20], [21], multiple-antenna architectures [22], [23], [24], [25], cognitive radio and capacity overlays [26], [27], [28] and power control [14], [29].

A common shortcoming of the transmission capacity approach is that it considers a typical network snapshot, and hence only simultaneous single hop transmissions. The primary goal of this paper is to address this limitation, and we develop a related metric which we term the *random access transport capacity* because it presumes packets are transported end-to-end over some distance $R$, but assumes independent locations and transmissions for the interferers. In that sense, one contribution of the paper is to find a middle ground between the transport and transmission capacity approaches into the capacity of multihop wireless networks. As one might expect, the results of this paper follow the $\Theta(\sqrt{n})$ scaling law, but provide additional information on the preconstants.

Complementary to the transport capacity research, some recent work has formulated the multihop capacity problem as a line network without additional network interference [30], [31]. Both of these papers agree that numerous hops are helpful only in the "power-limited" regime, that is where the spectral efficiency is low and overcoming noise is the primary concern. Both also find that in the "bandwidth-limited regime" – when the SNR is high – that additional hops decrease the end-to-end throughput due to the use of extra time-slots. Two notable theoretical results are $i)$ that the end-to-end capacity scales as $O(\log M)$ for $M$ hops [30] and $ii)$ that by using optimal time-sharing on each hop $m = 1, \dots, M$ to achieve the per-hop capacity $c_m$, the end-to-end capacity with no interference increases from $C_{\text{no int}} = \min_m(c_m)$ in [30] to

$$C_{\text{no int}} = \left( \sum_{m=1}^{M} \frac{1}{c_m} \right)^{-1} \tag{1}$$

as derived in [31]. Intuitively, (1) tells us that each additional hop lowers the capacity unless the hop capacities $c_m$ increase by a sufficient amount (due to higher SNR). Hence this equation also captures a basic tradeoff between higher per hop capacity (more short hops) and the desire for fewer total dimensions or transmission (fewer, and hence longer, hops). This tradeoff has been considered from many possible perspectives in [32], but in the present paper we are able to quantify it exactly in terms of a transport capacity-like metric, and with interference.

Finally, we note that a recent paper [33] has considered multihop capacity in a Poisson field of interference. The key distinction in this work is that we focus on achievable end-to-end throughput and optimal transmission strategies and hop count rather than the stability of queues. As [33] argues, the delay-optimizing number of hops appears to also be throughput-optimizing, so our conclusions are not contradictory. We also consider noise whereas they do not, which is quite important in the power-limited regime where multihop is beneficial.

### B. Contributions and Organization

In this paper we develop a new, quite general model for end-to-end throughput in a multihop wireless network. We term the resulting metric *random access transport capacity* since the analysis requires all transmissions to be independent, which precludes cooperative transmission scheduling among the nodes, since this would generally couple transmissions and the active transmitter locations would no longer be independent. Note, however, that the model does not preclude cooperative or multipacket reception, although we do not consider such approaches in this paper. The general model includes arbitrary paths of $M$ hops and an end-to-end delay/energy constraint quantified as the total allowable number of transmissions per packet, $A$. A closed-form upper bound on random access transport capacity is obtained by making the $M$ hops equidistant and by letting $A \to \infty$. This upper bound is tight for moderate values of $A$.

From the upper bound, the optimal number of hops $M^*$ can be computed as a function of the network parameters – in fact this was a prerequisite to finding the upper bound. For example, we can show that $M^* \propto R\beta^{\frac{1}{\alpha}}\sqrt{\lambda}$, where $R$ is the end-to-end distance, $\beta$ is the per-hop SINR required for successful communication at the transmitted rate, $\alpha$ is the path loss exponent, and $\lambda$ is the spatial density of transmitters. Additionally, the optimal per-hop success probability (or equivalently, outage probability) and optimal average number of retransmissions per hop can be directly found. This exposes an optimal tradeoff between highly reliable transmissions – achieved with short hops (large $M$) and hence a low per-hop outage – and unreliable transmissions, which make up for the retransmissions by requiring fewer hops. The upper bound is shown to be accurate for the purposes of predicting the optimal number of hops $M^*$ and the optimal per-hop success (or outage) probability.

An advantage of the novel framework developed in this paper is that the end-to-end throughput (here, the random access transport capacity) uses a small, finite, and computable number of hops, and results are given as a function of the $\mathcal{S}-\mathcal{D}$ separation $R$. We prove that the random access transport capacity follows the square-root scaling law, while providing easily computable preconstants. We expect that this approach can be extended to better understand and compare many specific candidate technologies and design approaches in multihop wireless networks.

The rest of this paper is organized as follows. First in Section II, we define the overarching network model and metrics used in this paper. We then develop and derive the upper bound in Section III, which is the main result of the paper. Implications and observations based on the upper bound derivation are discussed in Section IV. Numerical results are given in Section V, in which the upper bound is evaluated against a numerical exact solution (for finite $A$) and the above insights from the results are discussed.

## II. Network Model and Metrics

We consider a large wireless ad hoc network where a typical source node $\mathcal{S}$ drawn from a homogenous 2-D Poisson point process (PPP) of intensity $\lambda$ wishes to transmit to a destination node $\mathcal{D}$ that is a distance $R$ away in a random direction and is not part of the PPP. All transmitters (desired and interfering) have fixed transmit power $\rho$, which in order to simplify notation is more precisely the average radiated power at a distance of 1 meter from the transmitter[1]. The noise power is $\eta$ and the channel strength is determined by path loss and fading, so the received power at a distance $R$ is $\rho\chi R^{-\alpha}$, where $\alpha > 2$ is the path loss exponent and $\chi$ results from iid Rayleigh fading, which means $\chi \sim \exp(1)$. The SNR is defined in this paper as $\mathrm{SNR} := \rho R^{-\alpha}/\eta$ regardless of the number of hops taken, so dividing the route into hops improves the per-hop SNR, as it should.

### A. Single Hop, Single Transmission

The simplest case is to consider a single transmission over the entire $\mathcal{S} - \mathcal{D}$ distance $R$. In this case, the received SINR, where the reference receiver is defined to be at the origin, is

$$\mathrm{SINR} = \frac{\rho\chi_0 R^{-\alpha}}{\sum_{i\in\Pi}\rho\chi_i|X_i|^{-\alpha} + \eta}. \tag{2}$$

The interfering node locations in (2) are drawn from a stationary (Poisson point process (PPP) on the plane of intensity $\lambda$, denoted $\Pi = \{X_i\}$, where each $X_i \in \mathbb{R}^2$ is the location of an interfering transmitter. In the single-hop model, all interferers are sources, themselves. This is a realistic model assuming the transmitting nodes in the network are randomly and independently located and do not cooperate.

Under this model, the probability of success for sending a packet from source to destination can be found to be

$$p_s = p_s(1) = \exp\left\{-\frac{\beta\eta}{\rho R^{-\alpha}} - \lambda\beta^{\frac{2}{\alpha}}K_\alpha R^2\right\}, \tag{3}$$

[1]The authors have proven in [14] and [29] that pairwise power control does not have a significant effect on outage probability or transmission capacity. Path loss models that avoid the singularity for $R \to 0$ are discussed in [34].

where $K_\alpha = 2\pi^2/(\alpha \sin(2\pi/\alpha))$. Obtaining this expression is crucial to the results in this paper but nontrivial: for the derivation the reader is referred to [35]. Note that a similar exact result exists for Nakagami fading [22] as well as a tight bound on success probability with path loss only (no fading) [13], but we will use the expression for Rayleigh fading in (3) throughout this paper.

This single hop model has been fairly well-studied in recent years. For such a model, we define the *random access transport capacity*, in this case for single hop, to be

$$C^{\mathrm{sh}} = p_s \lambda \log(1+\beta) R. \tag{4}$$

In words, $C^{\mathrm{sh}}$ gives the density of successful transmissions at rate $\log(1+\beta)$ that can span a $\mathcal{S}-\mathcal{D}$ distance of $R$: therefore $C$ has units of bps/Hz/m. As $R$ increases, $p_s \to 0$ exponentially fast per (3). If $p_s$ is fixed to be $1-\epsilon$ for some maximum outage probability $\epsilon$, the $R$ term is dropped in (4), and (3) is inverted to find $\lambda$, then (4) gives the so-called transmission capacity. The goal in this paper is to move beyond the single-hop, single-transmission model to a network that allows multiple hops and multiple transmissions per hop, while retaining some of the tractability of the transmission capacity model.

### *B. Multiple Hops, Multiple Transmissions per Hop*

Now, allow the $\mathcal{S}-\mathcal{D}$ distance $R$ to be subdivided into $M$ hops having not necessarily equal distances $r_m$, where $R = \sum_{m=1}^{M} r_m$. On each hop, retransmissions are allowed and acknowledgments are used. Therefore, the number of transmissions on each hop $m$ is denoted by the geometric random variables $(T_1(M), \ldots, T_M(M))$, where each $T_m(M) \in \{1, 2, \ldots\}$, and the total number of transmissions required to move a packet from $\mathcal{S}$ to $\mathcal{D}$ is

$$T(M) = \sum_{m=1}^{M} T_m(M). \tag{5}$$

The notion of outage comes in naturally by constraining the total number of transmissions per packet, which is functionally a constraint on the total amount of delay and energy per packet, for example due to a timeout. If the total transmission per packet allowance is $A$ then $T(M) \leq A$ is required for successful transmission, so an outage event occurs when $T(M) > A$. Therefore, the total number of *actual* transmissions per packet is $\min(T(M), A)$, or more compactly, $T(M) \wedge A$. We assume that each $\mathcal{S}-\mathcal{D}$ pair only transmits a single packet at a time along the entire multihop path: in other words there is no intra-route spatial reuse[2]. The interference density is still $\lambda$ as in the single-hop case, but each packet is sent $T(M) \wedge A$ times, causing the effective rate per $\mathcal{S}-\mathcal{D}$ pair to be $\log(1+\beta)/(T(M) \wedge A)$ to account for the retransmissions. In other words, if the transmitted bit rate at $\mathcal{S}$ is $\log(1+\beta)$, because $\mathcal{S}$ has to wait $T(M) \wedge A$ packets before transmitting again, the effective $\mathcal{S}-\mathcal{D}$ data rate is lowered by this factor. This also presumes that $\mathcal{S}$ is instantaneously notified when the packet is received successfully at $\mathcal{D}$ so it can send the next packet.

*Definition 1: Random access transport capacity.* The random access transport capacity $C$ for a multihop wireless network is the maximum average source to destination rate that can be sustained reliably over a distance $R$ with at most $A$ transmission attempts per packet, normalized by the area of the network. Formally, it can be characterized by maximizing over the number of hops as:

$$C(A) = \max_{M \in \{1,\ldots,A\}} \mathbb{P}(T(M) \leq A) \cdot \lambda \cdot \frac{\log(1+\beta)}{\mathbb{E}[T(M) \wedge A]} R \tag{6}$$

$$= \lambda \log(1+\beta) R \max_{M \in \{1,\ldots,A\}} \frac{\mathbb{P}(T(M) \leq A)}{\mathbb{E}[T(M) \wedge A]} \quad \text{(bps/Hz/m)}. \tag{7}$$

This quantity therefore determines the end-to-end rate that can be supported when the source nodes have density $\lambda$. The challenge is in computing the probability of success $\mathbb{P}(T(M) \leq A)$ and in computing the expectation of a minimum, *i.e.,* $\mathbb{E}[T(M) \wedge A]$. In the general model thus espoused, $T$ is the sum of $M$ independent but not identical random variables $(T_1(M), \ldots, T_M(M))$, where $T_m(M) \sim \text{Geo}(p_s(M))$, for $p_s(M)$ to be defined in (9). For each transmission per hop to have independent success probability requires sufficient diversity in the interference and signal strength per transmission attempt, which could possibly be achieved through diversity techniques such as frequency hopping. Even with these simplifications, the general model does not appear to be tractable. In the next section, we develop an upper bound on the random access transport

[2]This is an important simplification, as it avoids dependent simultaneous transmissions along a path, as well as difficulties in accounting for non-Poisson interference. From a network-wide perspective the lack of spatial reuse for a given path does not affect the transport capacity because other $\mathcal{S}-\mathcal{D}$ pairs can utilize the space instead. Strictly speaking, the relays must be deterministically placed on the line between $\mathcal{S}$ and $\mathcal{D}$ rather than drawn from $\Pi$ in order to maintain a PPP for the interference. Since exactly one node transmits per route at a time, the interferers in each time slot form a PPP in this case. We also note that the spatio-temporal correlations in a Poisson field of interference have recently been studied in [36]: here, the effect would be to increase the independence as $M$ increases.

capacity that is computable in closed-form.

## III. Upper Bound: Guaranteed End-to-End Delivery, No Delay-Energy Constraint

### A. Establishing an upper bound

To develop an upper bound on the random access transport capacity, we make two important simplifications. First, we assume the hop lengths are all equidistant, that is $r_m = R/M$, for $m = 1, \ldots, M$. This is best-case because $p_s$ in (3) is $o(e^{-r_m^2})$, so the increase in $p_s$ on short hops is outweighed by the decrease in $p_s$ on longer hops. In other words, any perturbation from equal-length hops increases the end-to-end outage probability, all else being equal.

This assumption allows much improved tractability because now the probability of success is the same for all hops, so the number of transmissions $T_m(M)$ required on a given hop for success becomes iid geometric with parameter $p_s(M)$ to be defined in (9). The second important simplification is that we relax the delay-energy constraint. Formally, we let $A \to \infty$ in $\mathbb{P}(T(M) \leq A)/\mathbb{E}[T(M) \wedge A]$. This simplifies both the numerator and denominator: the probability of end-to-end success $\mathbb{P}(T(M) \leq A)$ tends to 1, and $\mathbb{E}[T(M) \wedge A]$ simplifies to $\mathbb{E}[T(M)]$. These considerations yield the following Lemma and Corollary.

*Lemma 1:* For all $A \in \mathbb{Z}_+$ and all $M \in \{1, \ldots, A\}$:

$$\frac{\mathbb{P}(T(M) \leq A)}{\mathbb{E}[T(M) \wedge A]} \leq \frac{1}{\mathbb{E}[T(M)]}. \tag{8}$$

*Proof:* See Appendix. ■

Although the numerator $\mathbb{P}(T(M) \leq A) \to 1$ and the denominator $\mathbb{E}[T(M) \wedge A] \to \mathbb{E}[T(M)]$ both increase, the numerator does so slightly more quickly, resulting in an upper bound. Under the upper bound assumption of equally spaced hops ($r_m = R/M$), the per-hop probability of success becomes:

$$p_s(M) := \exp\left\{-\frac{\beta}{\mathrm{SNR}} M^{-\alpha} - \lambda R^2 \beta^{\frac{2}{\alpha}} K_\alpha M^{-2}\right\}, \tag{9}$$

and the probability mass function of the number of transmission attempts $T_m(M)$ on each hop $m$ is geometric with probability of success $p_s(M)$:

$$P(T_m(M) = t) = (1 - p_s(M))^{t-1} p_s(M). \tag{10}$$

The average number of attempts per hop is therefore $\mathbb{E}[T_m(M)] = 1/p_s(M)$. Because the $M$ hops are iid it follows that the expected number of total transmissions required to move a packet from source to destination is $\mathbb{E}[T(M)] = M\mathbb{E}[T_m(M)] = M/p_s(M)$. Combining these observations yields the following upper bound on the capacity.

*Corollary 1:* An upper bound on the capacity (7) is

$$C^{\text{ub}}(A) = \lambda \log(1+\beta) R \max_{M \in \{1,\ldots,A\}} \frac{p_s(M)}{M}. \tag{11}$$

Note that although we relax the $A$ constraint in Lemma 1 *inside* the optimization in (7) to establish the upper bound, we retain a finite $A$ in the *range* of $M \in \{1, \ldots, A\}$.

### B. Optimal number of hops

The next step is to determine asymptotic optimal number of hops $M^*$ that maximizes $C^{\text{ub}}(A)$ as $A \to \infty$. This is given by the following theorem and corollaries, which give a positive real value for $M^*$ although the actual quantity would necessarily be an integer. We leave it as a continuous quantity in this paper for tractability and generality. Once $M^*$ is computed, the optimal average number of transmissions required per hop can be readily determined as $\mathbb{E}[T_m(M^*)] = p_s^{-1}(M^*)$.

*Theorem 1: Optimal number of hops, $M^*$.* The number of hops $M^*$ that optimizes the asymptotic transport capacity density upper bound $\lim_{A\to\infty} C^{\text{ub}}(A)$, *i.e.,*

$$M^* := \arg \max_{M=1,2,\ldots} \frac{p_s(M)}{M} \tag{12}$$

is the solution to the equation

$$M^\alpha - 2\lambda\beta^{2/\alpha} K_\alpha R^2 M^{\alpha-2} - \frac{\beta\eta R^\alpha}{\rho}\alpha = 0. \tag{13}$$

This results in closed-form expressions[3] for $M$ only when $\alpha \in \{3, 4, 6, 8\}$, in which case $M^*$ is the largest positive root of (13).

*Proof:* The objective is

$$\frac{p_s(M)}{M} = \frac{1}{M} \exp(-k_1 M^{-\alpha} - k_2 M^{-2}) \tag{14}$$

where $k_1 = \frac{\beta\eta}{\rho R^{-\alpha}} = \frac{\beta}{\text{SNR}}$ and $k_2 = \lambda\beta^{2/\alpha} K_\alpha R^2$. Taking the derivative and setting equal to zero eventually gives

$$\exp(k_1 M^{-\alpha} + k_2 M^{-2}) \left(1 - k_1 \alpha M^{-\alpha} - 2k_2 M^{-2}\right) = 0 \tag{15}$$

[3]By closed-form, we mean direct computation is possible using only basic arithmetic operations, simple trigonometric functions, and radicals.

which gives that

$$1 - k_1 \alpha M^{-\alpha} - 2k_2 M^{-2} = 0 \tag{16}$$

or in polynomial form

$$M^{\alpha} - 2k_2 M^{\alpha-2} - k_1 \alpha = 0. \tag{17}$$

By the Abel-Ruffini theorem, a formula solution to a polynomial equation only exists for when the degree of the polynomial is 4 or less. The path loss exponent $\alpha$ is of physical interest[4] only when $1.5 \lessapprox \alpha \lessapprox 6$. In this range, $M$ can be found in closed-form only for $\alpha \in \{2, 3, 4, 6\}$, although it can also be found in principle for $\alpha \in \{1, 8\}$. The solutions for $\alpha = 6$ and $\alpha = 8$ would follow similarly to the $\alpha = 3$ and $\alpha = 4$ solutions.

■

The corollaries for $\alpha = 3$ and $\alpha = 4$ follow immediately. A solution in terms of $K_\alpha$ can be found for $\alpha = 2$, but for $\alpha \to 2, K_\alpha \to \infty$ so this result does not exist. Intuitively, when $\alpha \leq 2$ the expected interference at any point in the network is infinite since the expected number of interferers increases as $d^2$ but the interference fall off is $d^\alpha$. Thus, the optimal number of hops for $\alpha = 2$ is technically infinite. In the sequel we write $M^* = M^*(\alpha)$ to emphasize the sensitivity of the optimal number of hops on the path loss exponent.

*Corollary 2:* Solving (13) with $\alpha$ = 3 yields two possible solutions, depending on the polarity of the equation's discriminant $D = \frac{9k_1^2}{4} - \frac{8k_2^3}{27}$. If $D \geq 0$,

$$M^*(3) = \beta^{\frac{1}{3}} R \left[ \sqrt[3]{\frac{3\eta}{2\rho} + f} + \sqrt[3]{\frac{3\eta}{2\rho} - f} \right], \tag{18}$$

where

$$f = \sqrt{\left(\frac{3\eta}{2\rho}\right)^2 - \frac{8K_3^3}{27}\lambda^3}, \tag{19}$$

and $K_3 = \frac{4\sqrt{3}}{9}\pi^2 \approx 7.6$. When $D < 0$

$$M^*(3) = \frac{2\sqrt{2\lambda K_3}}{3^{\frac{1}{6}}} \beta^{\frac{1}{3}} R \cos\left( \frac{1}{3} \text{arc}\cos\left[ \frac{3\sqrt{3}\eta}{4\sqrt{2}\rho(\lambda K_3)^{\frac{3}{2}}} \right] \right) \tag{20}$$

*Proof:* See Appendix. ■

[4]That is, there exist empirical studies that have found $\alpha$ to be in this range

Although the expressions (18) and (20) at first appear to be quite different, in fact they have a quite similar dependence in terms of the main parameters of interest. Finally, we have the following corollary for $\alpha = 4$.

*Corollary 3:* Solving (13) with $\alpha = 4$ yields a single maximum positive, real solution for $M^*$ which is

$$M^*(4) = \beta^{\frac{1}{4}} R \sqrt{\lambda \frac{\pi^2}{2} + \sqrt{\lambda^2 \frac{\pi^4}{4} + \frac{4\eta}{\rho}}} \tag{21}$$

since $K_4 = \pi^2/2$.

*Proof:* See Appendix. ■

## C. The upper bound, $C^{\text{ub}}$

The hop-optimized random access transport capacity can now be given in closed-form for any choice of system parameters since

$$C^{\text{ub}} = \frac{\lambda \log(1+\beta) R}{\mathbb{E}[T_m(M^*)] \cdot M^*} = \lambda \log(1+\beta) R \frac{p_s(M^*)}{M^*}, \tag{22}$$

and we now have found exact expressions for $p_s(M^*)$ and $M^*$. Although a few assumptions are made to get to closed form – fixed equidistant relays, randomly located interferers, independent retransmissions, all transmissions at the same rate – this allows a simple closed-form end-to-end expression for multihop network throughput with both noise and interference.

For example, consider $\alpha = 4$, in which case at high SNR $M^* \to \beta^{\frac{1}{4}} R \pi \sqrt{\lambda}$, which yields with (9) the following quite simple exact expression for transport capacity

$$\lim_{\text{SNR} \to \infty} C^{\text{ub}*}\Big|_{\alpha=4} = \frac{\sqrt{\lambda} \log(1+\beta)}{\pi \beta^{\frac{1}{4}}} \exp\left(-\frac{\eta}{\rho \pi^4 \lambda^2} - \frac{1}{2}\right). \tag{23}$$

# IV. Upper Bound Implications

## A. Initial Insights from Theorem 1

It can be immediately observed that all the derived solutions for $M^*$ are precisely proportional to $R\beta^{\frac{1}{\alpha}}$. This makes sense: the optimal number of hops should scale linearly with the source-destination distance since this keeps the per-hop distance and hence the per-hop SNR and SINR constant for a certain $\rho, \eta, \lambda$. Similarly, since the received SINR is proportional to the received (desired) power $P_r$ over a distance $d$, we can see that

$$P_r = \rho d^{-\alpha} = \rho \left(\frac{M}{R}\right)^{\alpha}. \tag{24}$$

Since $P_r$ must scale linearly with $\beta$ with the interference ($\lambda$) held constant,

$$\rho \left(\frac{M}{R}\right)^{\alpha} = \text{constant} \cdot \beta \Rightarrow M \propto \beta^{\frac{1}{\alpha}}. \tag{25}$$

Less obviously, there is a trend in the expressions that $M \propto \sqrt{\lambda}$, which also makes intuitive sense. As the density increases, the interferers' relative distance (statistically) decreases as $\sqrt{\lambda}$, requiring a shorter communication distance by the same amount in order to maintain the same SINR. Hence, $R/M \propto \lambda^{-2}$ or $M \propto \sqrt{\lambda}$. This trend is more difficult to see in the $\alpha = 3$ case for $D < 0$ (20) but since

$$\cos\left(\frac{1}{3}\text{arc}\cos x\right) \approx \frac{\sqrt{3}}{2} + \left(1 - \frac{\sqrt{3}}{2}\right) x, \tag{26}$$

one of the two terms has $\sqrt{\lambda}$, the other having a $\lambda^{-1}$ term. As $\lambda$ increases, the $\lambda^{-1}$ would decrease leaving $M \propto \sqrt{\lambda}$.

In summary, from Theorem 1 and its corollaries, one can immediately determine the optimum number of hops for any plausible integer path loss exponent in terms of the salient network parameters. Non-integer values could be accurately estimated by standard interpolation techniques.

### B. Optimal success probability

Armed with the optimal number of hops $M^*$, it is straightforward to determine the optimum success probability per hop, $p_s(M^*)$. Interestingly, it can be expressed in two ways, where the first expression depends on the $\lambda$ but not SNR (neither noise nor interference power affect this expression), and for the second expression, vice versa.

*Lemma 2:* The optimal success probability per hop $p_s(M^*)$ as a function of $M^*$ is

$$p_s(M^*) = \exp\left\{\left(\frac{2}{\alpha} - 1\right)\frac{\lambda \beta^{\frac{2}{\alpha}} K_\alpha R^2}{(M^*)^2} - \frac{1}{\alpha}\right\} \tag{27}$$

$$= \exp\left\{\left(\frac{\alpha}{2} - 1\right)\frac{\beta}{\text{SNR}(M^*)^{\alpha}} - \frac{1}{2}\right\} \tag{28}$$

*Proof:* Since the optimal number of hops $M^*$ should satisfy (17), we can write the following two expressions

$$k_1 M^{-\alpha} = \frac{1 - 2k_2 M^{-2}}{\alpha}, \tag{29}$$

$$k_2 M^{-2} = \frac{1 - k_1 M^{-\alpha}}{2} \tag{30}$$

Plugging (29) into (9) gives (27), while plugging (30) into (9) gives (28). ■

The two expressions for $p_s(M^*)$ can be seen as the $p_s$ which optimally balances between interference (27) and noise (28). Note additionally that the optimal number of average retransmissions per hop is given by $T_m(M^*) = 1/p_s(M^*)$, which is the same for each hop.

### *C. Relation to general transport capacity: shared scaling law*

Although the primary merit of the approach of this paper is the ability to compute end-to-end rate results in closed-form, the results here also obey the well-known $\Theta(\sqrt{n})$ scaling law for transport capacity [1], [8], or in the lexicon of this paper, $\Theta(\sqrt{\lambda})$. This fact is formalized in the following theorem.

*Theorem 2: The random access transport capacity upper bound is* $\Theta(\sqrt{\lambda})$ *and can be stated formally as*

$$\lim_{\lambda \to \infty} \frac{C^{\text{ub}}}{\sqrt{\lambda}} = \frac{e^{-\frac{1}{2}} \log(1+\beta)}{\sqrt{2K_\alpha} \beta^{\frac{1}{\alpha}}}. \tag{31}$$

*Proof:* First, it can be shown that $M^* = c_0\sqrt{\lambda}$ for some constant $c_0$. Rearranging (17) gives

$$1 - 2\lambda \beta^{\frac{2}{\alpha}} K_\alpha R^2 M^{-2} = \frac{\beta\alpha}{\text{SNR} M^\alpha}. \tag{32}$$

For this equation to hold as $\lambda \to \infty$, clearly $M \to \infty$, which means that the quantity on the left of (32) must approach 0. Therefore, $M = \sqrt{\lambda} \cdot \sqrt{2K_\alpha} \beta^{\frac{1}{\alpha}} R$, *i.e.,* $c_0 = \sqrt{2K_\alpha}\beta^{\frac{1}{\alpha}} R$.

Second, from (28) it can be observed that $p_s \to \exp(-\frac{1}{2})$ as $\lambda \to \infty$, whereas (27) can be used to verify the value for $c_0$. Using (22) and the scaling results just derived gives

$$\lim_{\lambda \to \infty} C^{\text{ub}} = \lim_{\lambda \to \infty} \frac{\lambda p_s(M^*) \log(1+\beta) R}{M^*} \tag{33}$$

$$= \frac{\lambda \exp(-\frac{1}{2}) \log(1+\beta) R}{\sqrt{\lambda} c_0} \tag{34}$$

$$= \frac{\sqrt{\lambda} e^{-\frac{1}{2}} \log(1+\beta)}{\sqrt{2K_\alpha} \beta^{\frac{1}{\alpha}}} \tag{35}$$

■

This theorem suggests that the random access transport capacity even subject to the assumptions we have adopted in this paper is order optimal: it follows the square-root scaling law for source-destination transmissions in large wireless networks. That is, unscheduled, channel-blind transmissions can achieve – given well-positioned relays – a transport capacity that scales the same as optimally scheduled nearest neighbor routing. This complements a similar conclusion for slotted ALOHA in regular networks [12].

## V. Numerical Results and Insights

We set $R = 1$, $\beta = 3$, and as a default $\alpha = 3$, while the other parameters are varied. Note that the units of $R$ and $\lambda$ are arbitrary, but in a circle with $\mathcal{S}$ and $\mathcal{D}$ on opposite sides separated by a distance $R$, there would be on average $\lambda\pi(R/2)^2$ interferers. Because only together do the values of $\lambda$ and $R$ provide a description of the interference environment, $\lambda \gg 1$ may be reasonable to model a dense network or a long $\mathcal{S} - \mathcal{D}$ path when $R = 1$.

Fig. 1 demonstrates that the random access transport capacity upper bound is tight for a small and large number of allowed attempts $A$, but fairly loose in between. This is perhaps to be expected since the bound presumes $A \to \infty$. Figs 2 and 3 show the upper bound vs. the actual random access transport capacity (computed numerically) for $A = 6$ and $A = 12$, respectively, vs. the number of hops used, $M$. As would be expected by Fig. 1, the former is loose while the latter is fairly tight. What is interesting is that in both cases the maximizing $M$ is very similar. In other words, even when the capacity upper bound is loose, the value of $M^*$ that it predicts appears to be quite accurate. This is confirmed in Fig. 4: the $M^*$ from the UB and the actual $M^*$ are always within one hop of each other and asymptotically the same. Interestingly, the actual $M^*$ is not necessarily monotonic in $A$, as Fig. 4 demonstrates.

Now focusing on the upper bound, first consider the result of Theorem 1, which gave the optimum number of hops $M^*$. In Fig. 5 we observe that the $M^*$ is indeed $\Theta(\sqrt{\lambda})$ and that it obtains this scaling in the $\lambda \in (.1, 1)$ range over a wide range of SNR and for both $\alpha = 3$ and $\alpha = 4$. The scaling law here kicks in faster at high SNR, because at low SNR a higher contention density is required before the network becomes interference limited. Although the path loss exponent does not appear to affect the convergence, smaller path loss exponents require more hops to be taken in order to mitigate the cumulative interference. This may seem counter-intuitive because larger path loss exponents cause faster attenuation of the desired signal, which might seem to imply that more hops would be needed to maintain the signal strength, but this is not the case, even at very small interference densities.

In Fig. 6 we plot the random access transport capacity upper bound where we do not optimize for $M$, but rather see how it varies with the number of hops. The optimum throughput occurs at the integer value of $M$ closest to $M^*$, as expected. The random access transport capacity is considerably higher overall with larger $\lambda$ (*i.e.,* source density) and $\mathrm{SNR}$, as would be expected, and achieving that higher end-to-end throughput requires significantly more hops in order to maintain a high enough SINR on each hop. On the contrary, if a small number of hops was used in this case, despite the high SNR, the end-to-end throughput would be much lower than in the lightly loaded (small $\lambda$) case. Optimizing the number of hops – or at least getting close to $M^*$ – is very important.

Turning our attention to Theorem 2, we observe in Fig. 7 that the scaling law of $\Theta(\sqrt{\lambda})$ holds for large $\lambda$. Similar to Fig. 5 the scaling law kicks in faster at high SNR and the throughput increases with $\alpha$. Put optimistically, this means at low SNR (or low interference density) that the throughput actually increases quite a bit more quickly with $\lambda$ than the scaling law would suggest. Until the network is saturated with interference, more users can be added without having their throughput decrease.

## APPENDIX

## PROOF OF LEMMA 1

Fix a positive integer $A$ and $p \in [0,1]$. Define $T(M) = T_M \sim \mathrm{Pascal}(M,p)$ as the number of trials required for $M$ successes, and $S_A \sim \mathrm{Bin}(A,p)$ as the number of successes in $A$ trials. To prove the lemma it suffices to show $f(M) \geq 0$ for all $M = 0, \ldots, A$, where

$$f(M) := p\mathbb{E}[T_M \wedge A] - M\mathbb{P}(T_M \leq A). \tag{36}$$

The inequality is trivially true for $M > A$ and $M = 0$, and is straightforward to show for $M = 1$ and $M = A$. We first prove some key relationships for binomial random variables.

$$p\sum_{n=M}^{A}\mathbb{P}(S_n = M) = \mathbb{P}(S_{A+1} \geq M+1), \quad M = 0, \ldots, A-1 \tag{37}$$

$$\mathbb{P}(S_A \leq M-1) = (1-p)\mathbb{P}(S_{A-1} \leq M-1) + p\mathbb{P}(S_{A-1} \leq M-2) \tag{38}$$

$$(1-p)(M+1)\mathbb{P}(S_A = M+1) = p(A-M)\mathbb{P}(S_A = M) \tag{39}$$

To see (37), use the fact that $p\mathbb{P}(S_n = M) = \mathbb{P}(T_{M+1} = n+1)$ and $\{T_M \leq A\} = \{S_A \geq M\}$ to show:

$$p \sum_{n=M}^{A} \mathbb{P}(S_n = M) = \sum_{n=M}^{A} \mathbb{P}(T_{M+1} = n+1) = \mathbb{P}(T_{M+1} \leq A+1) = \mathbb{P}(S_{A+1} \geq M+1). \quad (40)$$

To see (38) simply observe the equality of the events, where $X_A$ indicates success on hop $A$:

$$\{S_A \leq M-1\} = \{(S_{A-1} \leq M-1) \cap (X_A = 0) \cup (S_{A-1} \leq M-2) \cap (X_A = 1)\}. \quad (41)$$

To see (39) use the "mixed product" identity: $(M+1)\binom{A}{M+1} = (A-M)\binom{A}{M}$. That is, with $D$ as the difference of the two sides in (39):

$$\begin{aligned} D &\equiv (1-p)(M+1)\mathbb{P}(S_A = M+1) - p(A-M)\mathbb{P}(S_A = M) \\ &= (1-p)(M+1)\binom{A}{M+1} p^{M+1}(1-p)^{A-M-1} - p(A-M)\binom{A}{M} p^M (1-p)^{A-M} \\ &= p^{M+1}(1-p)^{A-M}\left((M+1)\binom{A}{M+1} - (A-M)\binom{A}{M}\right) = 0 \end{aligned} \quad (42)$$

Using (37) we have the following development for $f(M)$:

$$\begin{aligned} f(M) &\equiv p\mathbb{E}[T_M \wedge A] - M\mathbb{P}(T_M \leq A) \\ &= p\left(\sum_{n=M}^{A} n\mathbb{P}(T_M = n) + A\mathbb{P}(T_M > A)\right) - M\mathbb{P}(T_M \leq A) \\ &= p^2 \sum_{n=M}^{A} n\mathbb{P}(S_{n-1} = M-1) + pA(1 - \mathbb{P}(T_M \leq A)) - M\mathbb{P}(T_M \leq A) \\ &= pA + Mp\sum_{n=M}^{A} \mathbb{P}(S_n = M) - (pA+M)\mathbb{P}(T_M \leq A) \\ &= pA + M\mathbb{P}(S_{A+1} \geq M+1) - (pA+M)\mathbb{P}(S_A \geq M) \\ &= (pA+M)\mathbb{P}(S_A \leq M-1) - M\mathbb{P}(S_{A+1} \leq M) \end{aligned} \quad (43)$$

We will prove $f(M) \geq 0$ by showing that $\Delta(M) \equiv f(M+1) - f(M) \geq 0$ for all $M = 0, \ldots, A-1$. Combined with $f(0) = 0$, this immediately yields that $f(M) \geq 0$ for all $M = 0, \ldots, A$, establishing the upper bound. Consider the difference $\Delta(M) \equiv f(M+1) - f(M)$ and

apply (38) twice:

$$\begin{aligned}
\Delta(M) &= [(pA+M+1)\mathbb{P}(S_A \le M) - (M+1)\mathbb{P}(S_{A+1} \le M+1)] - \\
&\quad [(pA+M)\mathbb{P}(S_A \le M-1) - M\mathbb{P}(S_{A+1} \le M)] \\
&= [(pA+M+1)\mathbb{P}(S_A \le M) - (M+1)\left((1-p)\mathbb{P}(S_A \le M+1) + p\mathbb{P}(S_A \le M)\right)] - \\
&\quad [(pA+M)\mathbb{P}(S_A \le M-1) - M\left((1-p)\mathbb{P}(S_A \le M) + p\mathbb{P}(S_A \le M-1)\right)] \\
&= \left((1-p)(2M+1) + pA\right)\mathbb{P}(S_A \le M) - \left((1-p)M + pA\right)\mathbb{P}(S_A \le M-1) - \\
&\quad \left((1-p)(M+1)\right)\mathbb{P}(S_A \le M+1) \qquad (44)
\end{aligned}$$

Now use the easy facts that:

$$\begin{aligned}
\mathbb{P}(S_A \le M-1) &= \mathbb{P}(S_A \le M) - \mathbb{P}(S_A = M) \\
\mathbb{P}(S_A \le M+1) &= \mathbb{P}(S_A \le M) + \mathbb{P}(S_A = M+1). \qquad (45)
\end{aligned}$$

Substitution of (45) gives:

$$\begin{aligned}
\Delta(M) &= \left((1-p)(2M+1) + pA\right)\mathbb{P}(S_A \le M) - \\
&\quad \left((1-p)M + pA\right)[\mathbb{P}(S_A \le M) - \mathbb{P}(S_A = M)] - \\
&\quad \left((1-p)(M+1)\right)[\mathbb{P}(S_A \le M) + \mathbb{P}(S_A = M+1)] \\
&= ((1-p)M + pA)\mathbb{P}(S_A = M) - (1-p)(M+1)\mathbb{P}(S_A = M+1) \qquad (46)
\end{aligned}$$

Apply (39) to the previous equation

$$\begin{aligned}
\Delta(M) &= ((1-p)M + pA)\mathbb{P}(S_A = M) - (1-p)(M+1)\mathbb{P}(S_A = M+1) \\
&= ((1-p)M + pA)\mathbb{P}(S_A = M) - (1-p)(M+1)\frac{p(A-M)}{(1-p)(M+1)}\mathbb{P}(S_A = M) \\
&= \mathbb{P}(S_A = M)(((1-p)M + pA) - p(A-M)) \\
&= \mathbb{P}(S_A = M)(M - pM + pA - pA + pM) \\
&= M\mathbb{P}(S_A = M) \ge 0 \qquad (47)
\end{aligned}$$

## Proof of Corollary 2

For the $D \ge 0$ case, a necessary condition on $\lambda$ and the transmit power can be formulated by inserting $k_1 = \frac{\beta\eta}{\rho R^{-\alpha}} = \frac{\beta}{\mathrm{SNR}}$ and $k_2 = \lambda\beta^{2/\alpha}K_\alpha R^2$ to get

$$\lambda \le \left(\frac{\eta}{\rho}\right)^{\frac{2}{3}}\left(\frac{3}{2}\right)^{\frac{5}{3}}\frac{1}{K_3} \Rightarrow \frac{\rho}{\eta} \le \sqrt{\left(\frac{3}{2}\right)^5 \frac{1}{K_3^3\lambda^3}}. \qquad (48)$$

The roots of such a 3rd order polynomial consist of 1 real and 2 complex roots. The real root is given by

$$M = \left(\frac{3k_1}{2} + \sqrt{D}\right)^{\frac{1}{3}} + \left(\frac{3k_1}{2} - \sqrt{D}\right)^{\frac{1}{3}} \tag{49}$$

which results in (18).

When $D < 0$ there exist three distinct real roots. The necessary conditions (48) are simply the converse of (48), but the polynomial solution requires some trigonometric calculations. Define

$$\zeta = \frac{2}{3}\sqrt{\frac{2k_2^3}{3}}, \quad \cos\phi = \frac{3k_1}{2\zeta} \Rightarrow \phi = \arccos\left(\frac{3k_1}{2\zeta}\right), \tag{50}$$

then the three possible solutions are

$$y_1 = 2\zeta^{\frac{1}{3}} \cos\left(\frac{\phi}{3}\right) \tag{51}$$

$$y_2 = 2\zeta^{\frac{1}{3}} \cos\left(\phi + \frac{2\pi}{3}\right) \tag{52}$$

$$y_3 = 2\zeta^{\frac{1}{3}} \cos\left(\phi + \frac{4\pi}{3}\right). \tag{53}$$

For these three real roots, Vieta's Theorem states that for a 3rd order polynomial $a_3y^3 + a_2y^2 + a_1y + a_0$ that the roots will satisfy $y_1y_2y_3 = -\frac{a_0}{a_3}$, which gives from (17) that

$$y_1y_2y_3 = 3k_1 \Rightarrow y_1y_2y_3 > 0. \tag{54}$$

Thus the roots must either be all positive or comprise one positive and two negative. They cannot all be positive since $y_1y_2 + y_1y_3 + y_2y_3 = 0$, also a consequence of Vieta's Theorem. Hence, there is a single positive root, which is $y_1$, and by expanding $\zeta$ and $\phi$ can be expressed as $M^* = y_1$, given in (20).

## PROOF OF COROLLARY 3

The polynomial for $\alpha = 4$ becomes

$$M^4 - 2k_2M^2 - 4k_1 = 0. \tag{55}$$

Using a dummy variable for $M^2$, this can be easily solved using the quadratic formula to give

$$M^*(4) = \pm\beta^{\frac{1}{4}}R\sqrt{\lambda\frac{\pi^2}{2} \pm \sqrt{\lambda^2\frac{\pi^4}{4} + \frac{4\eta}{\rho}}}, \tag{56}$$

of which only the solution given in (21) is both positive and real.

TABLE I

SUMMARY OF NOTATION AND PARAMETERS

| | |
|---|---|
| $\lambda$ | interference density, intensity of PPP $\Pi$ |
| $\alpha$ | path loss exponent ($\alpha > 2$) |
| $K_\alpha$ | A constant, $K_\alpha = 2\pi^2/(\alpha \sin(2\pi/\alpha))$ |
| $\beta$ | target (required) SINR per hop |
| $R$ | $\mathcal{S} - \mathcal{D}$ transmit distance |
| $r_m$ | transmit distance on hop $m$ |
| $\rho$ | transmit power |
| $\eta$ | noise power |
| SNR | $\mathcal{S} - \mathcal{D}$ end-to-end signal to noise ratio, $\mathrm{SNR} := \rho R^{-\alpha}/\eta$ |
| $p_s(M)$ | probability of success over a single hop |
| $M$ | number of hops (number of relay nodes is $M - 1$) |
| $T_m(M)$ | transmissions used on hop $m$ to achieve a successful transmission |
| $T(M)$ | total number of transmissions required per packet, $T(M) = \sum T_m(M)$ |
| $A$ | allowance on total number of transmissions per packet, $T(M) \leq A$ |
| $P_{\mathrm{out}}$ | probability of end-to-end outage, $P_{\mathrm{out}} = \mathbb{P}(T(M) > A)$ |
| $C(A), C^{\mathrm{ub}}(A)$ | random access transport capacity, and upper bound |

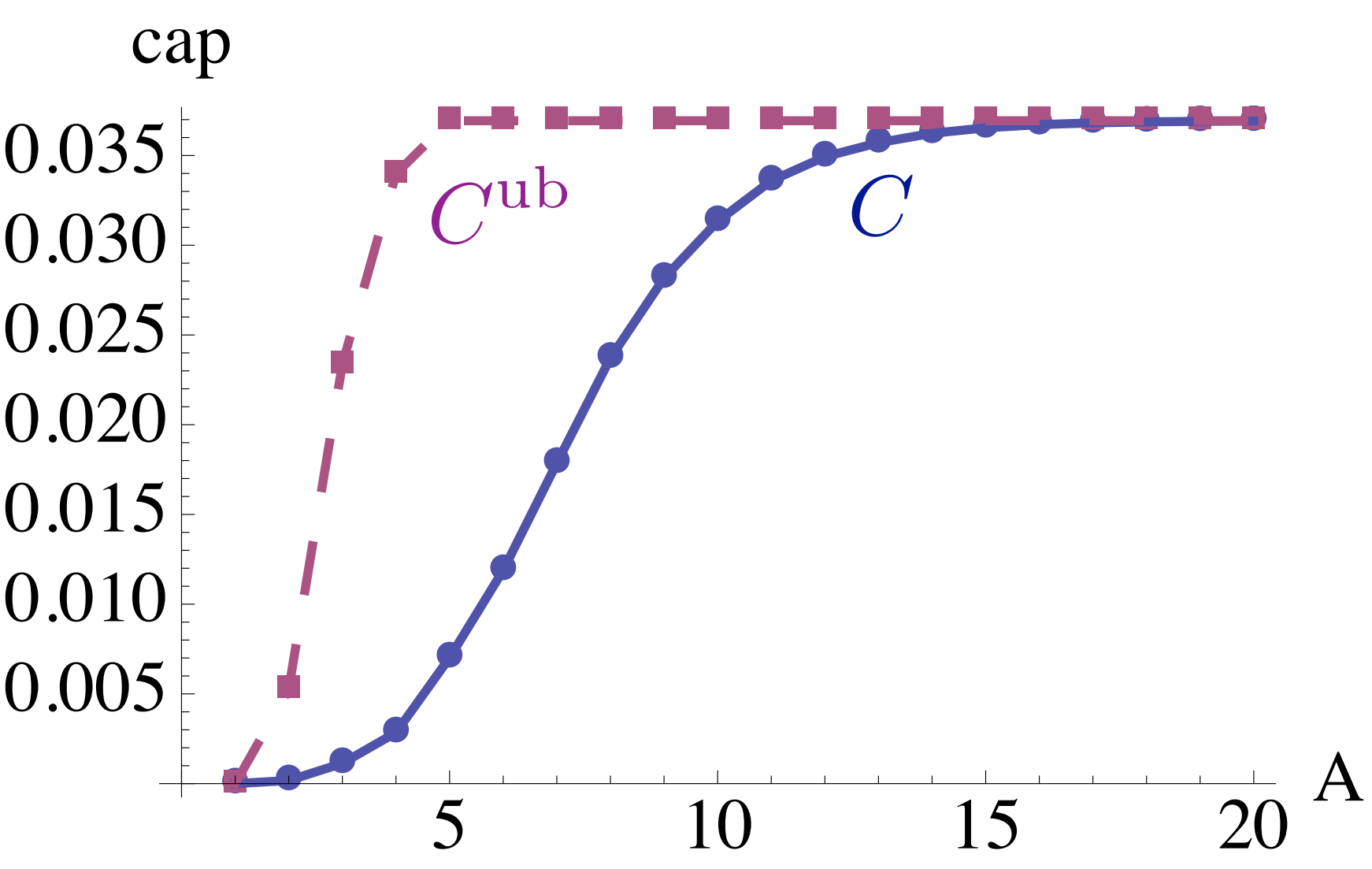

Fig. 1. The capacity $C(A)$ and its upper bound $C^{\rm ub}(A)$ versus the allowed number of transmission attempts $A$.

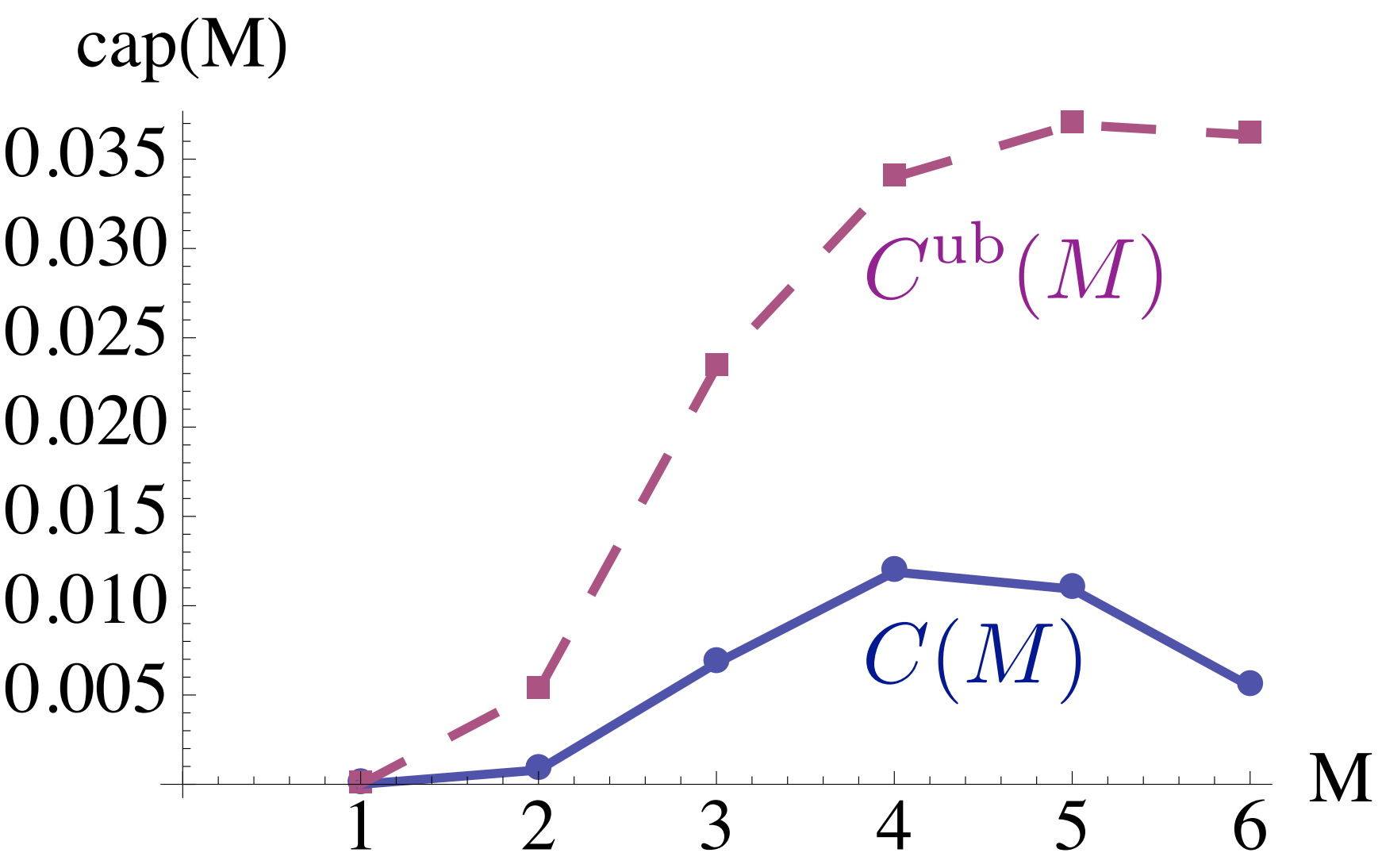

Fig. 2. The "capacity" and its upper bound versus the number of hops $M$ for each $M = 1, \ldots, A$ when $A = 6$. Note the maximizing $M^*$ for both curves are close to one another.

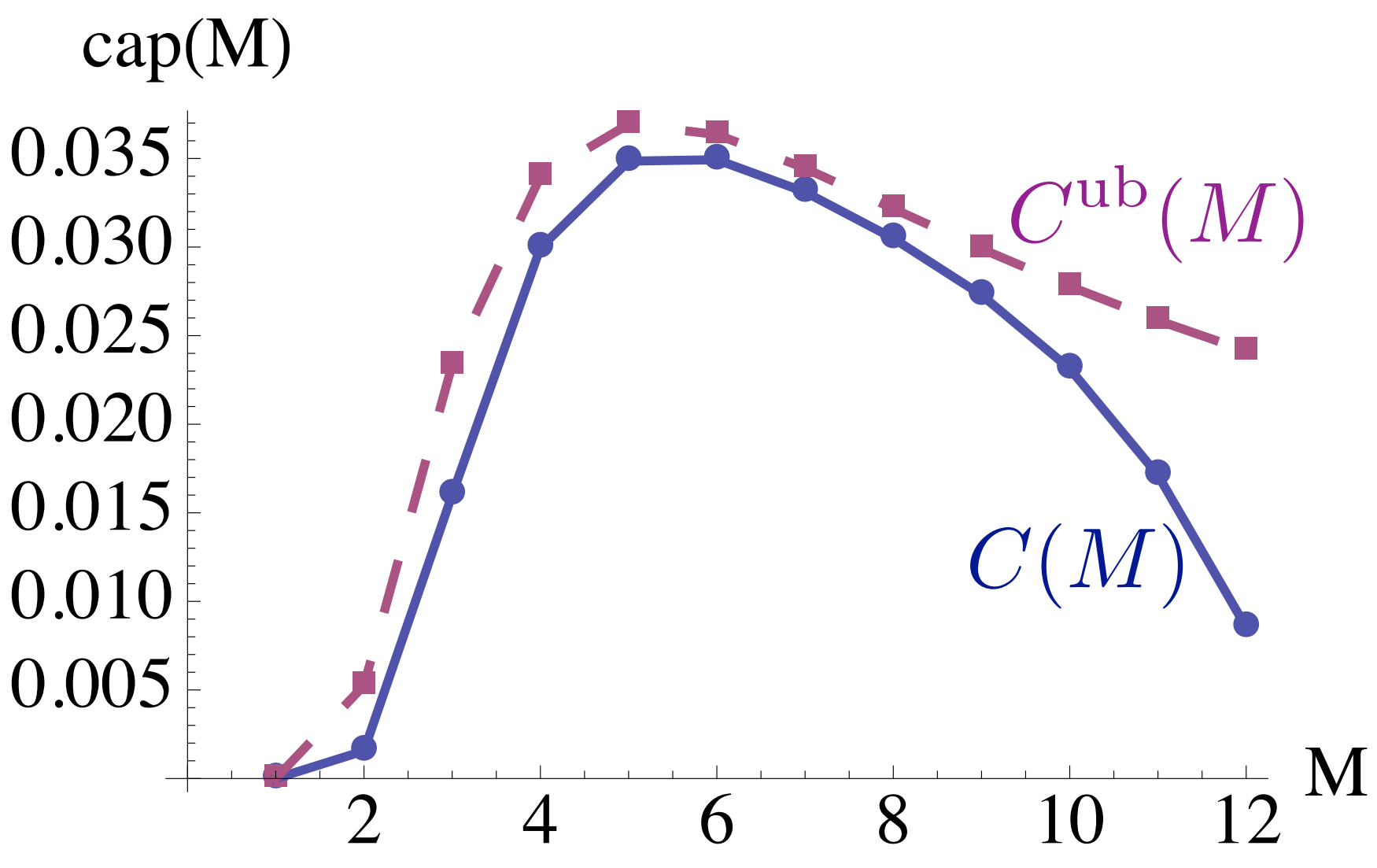

Fig. 3. The random access transport capacity and its upper bound versus the number of hops $M$ for each $M = 1, \ldots, A$ when $A = 12$. Note the maximizing $M^* = 5$ for both curves in this case.

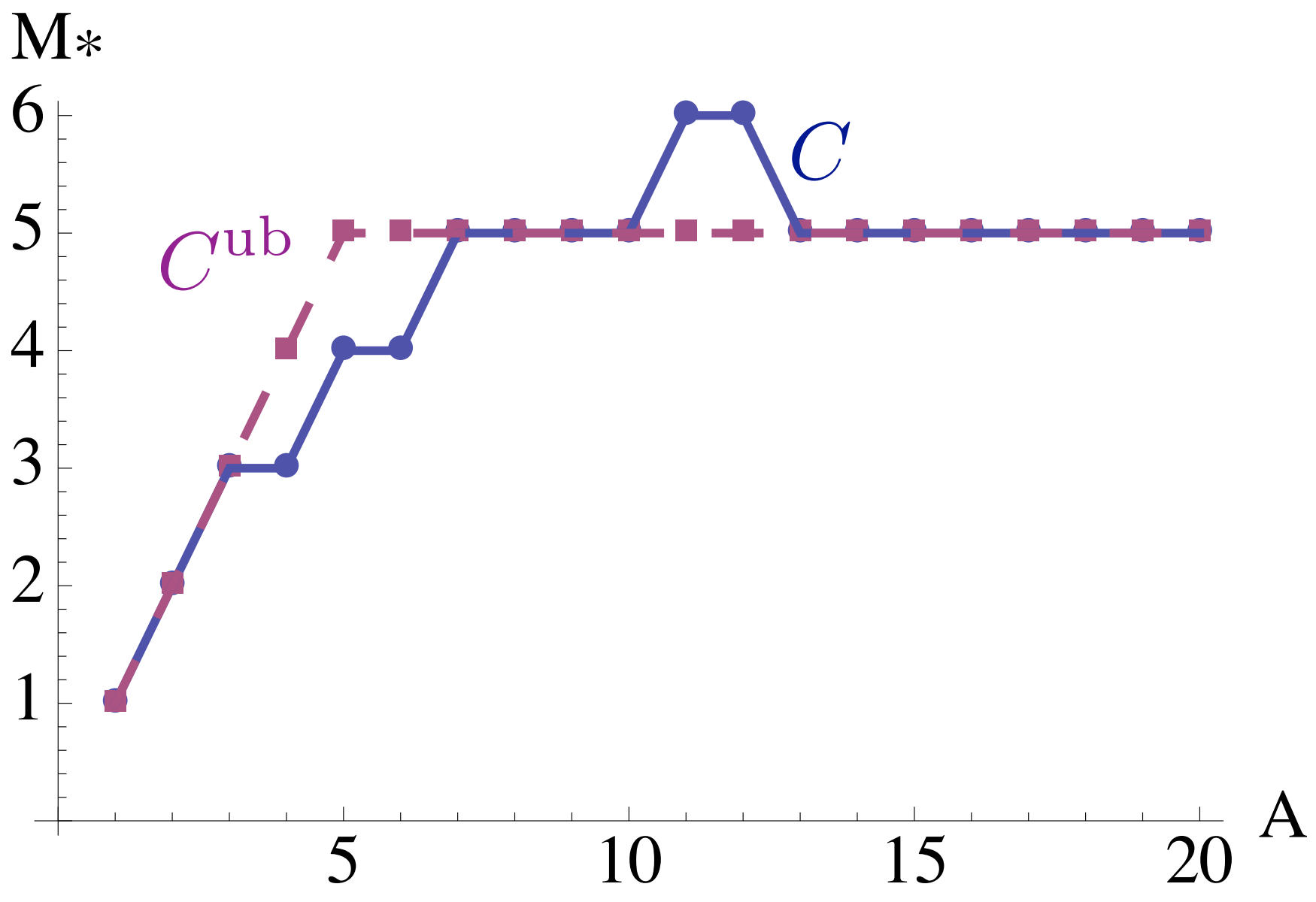

Fig. 4. The optimal number of hops $M^*(A)$ that maximizes capacity $C(A)$ and the upper bound $C^{\mathrm{ub}}(A)$ versus the allowed number of transmission attempts $A$.

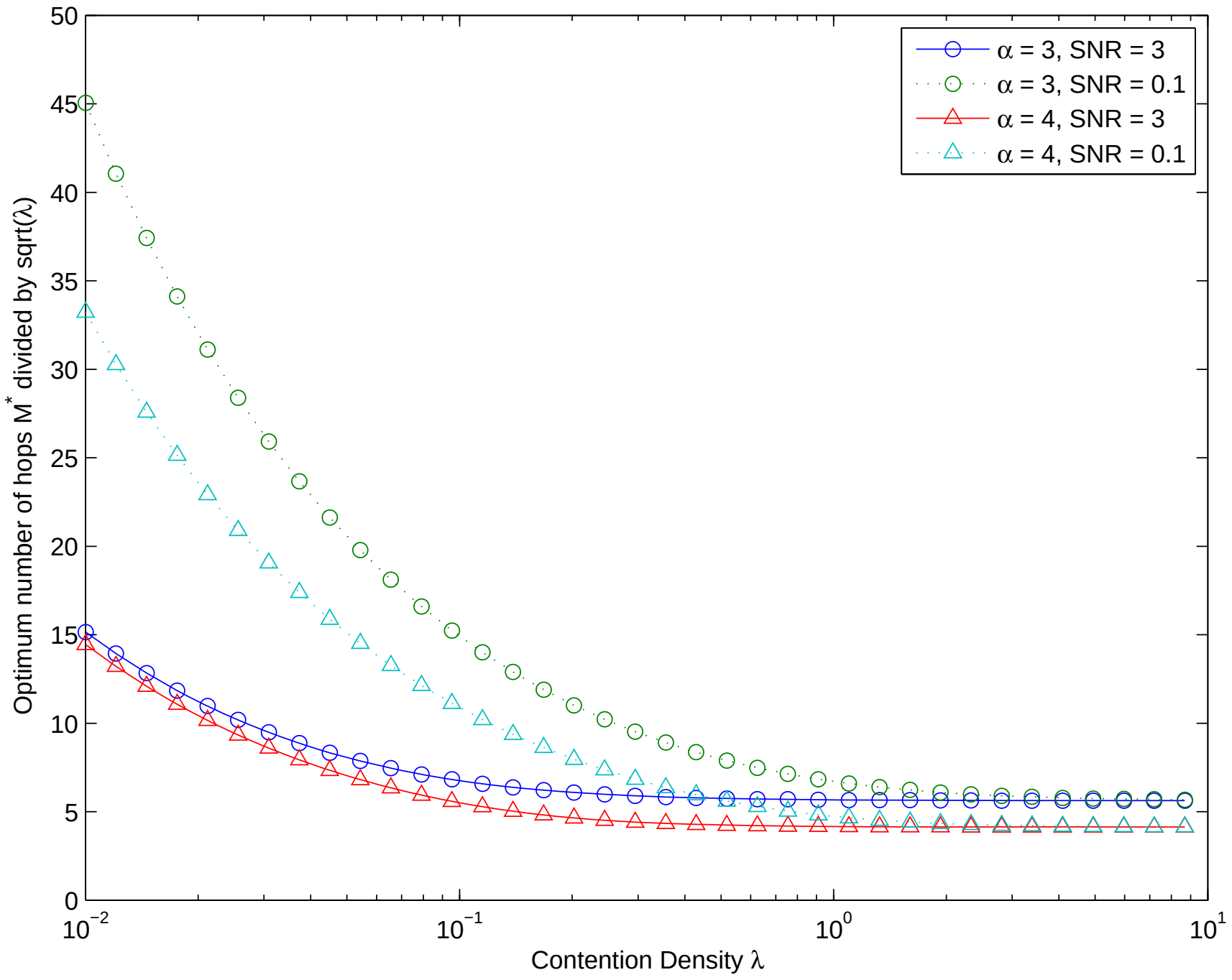

Fig. 5. The optimal number of hops $M^*$ normalized by $\sqrt{\lambda}$ vs. contention density $\lambda$. It can be observed that $M^* = \Theta(\sqrt{\lambda})$ as expected, with faster convergence at high SNR.

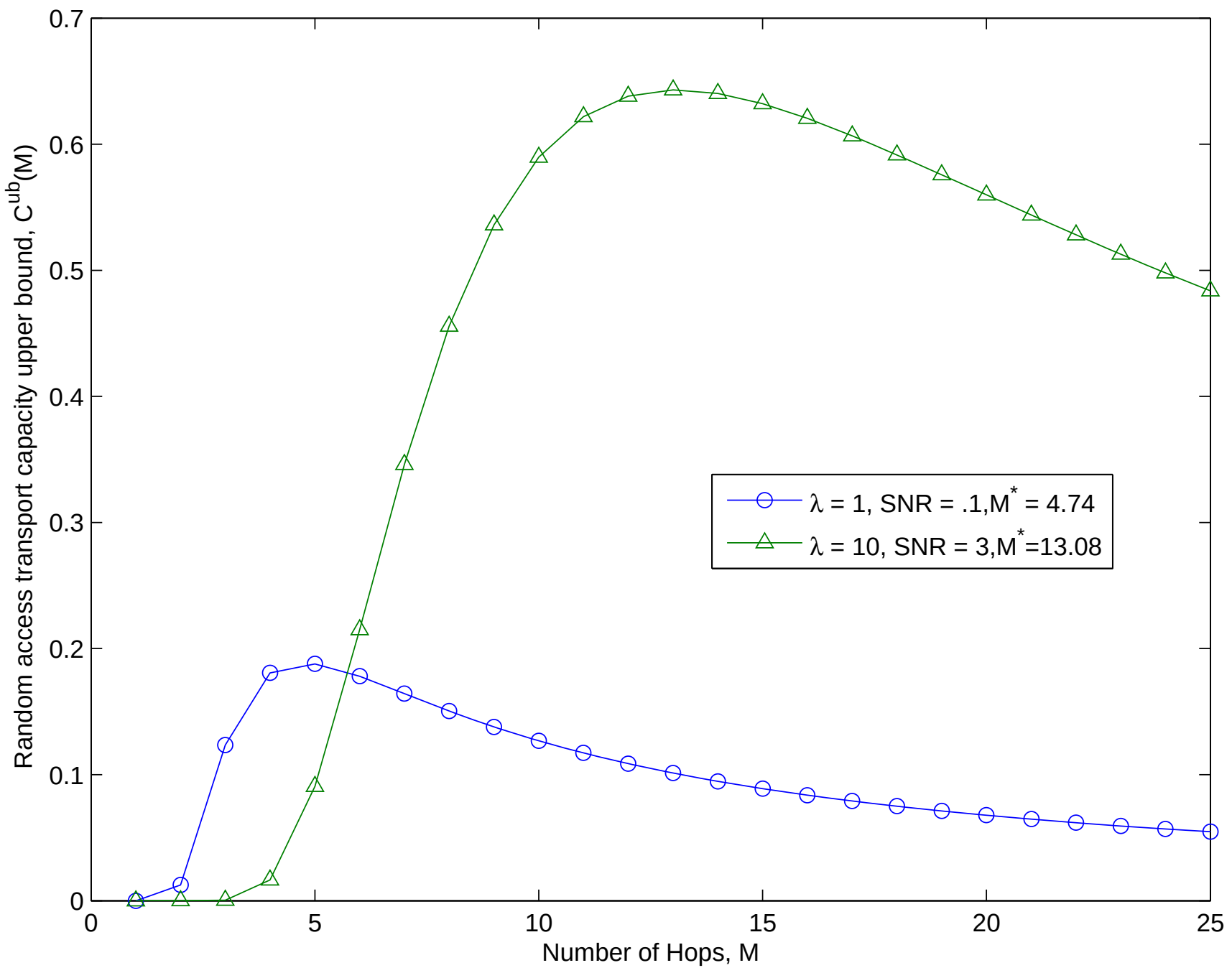

Fig. 6. The random access transport capacity upper bound is plotted vs. the number of hops $M$. There is a clear optimum near the derived $M^*$.

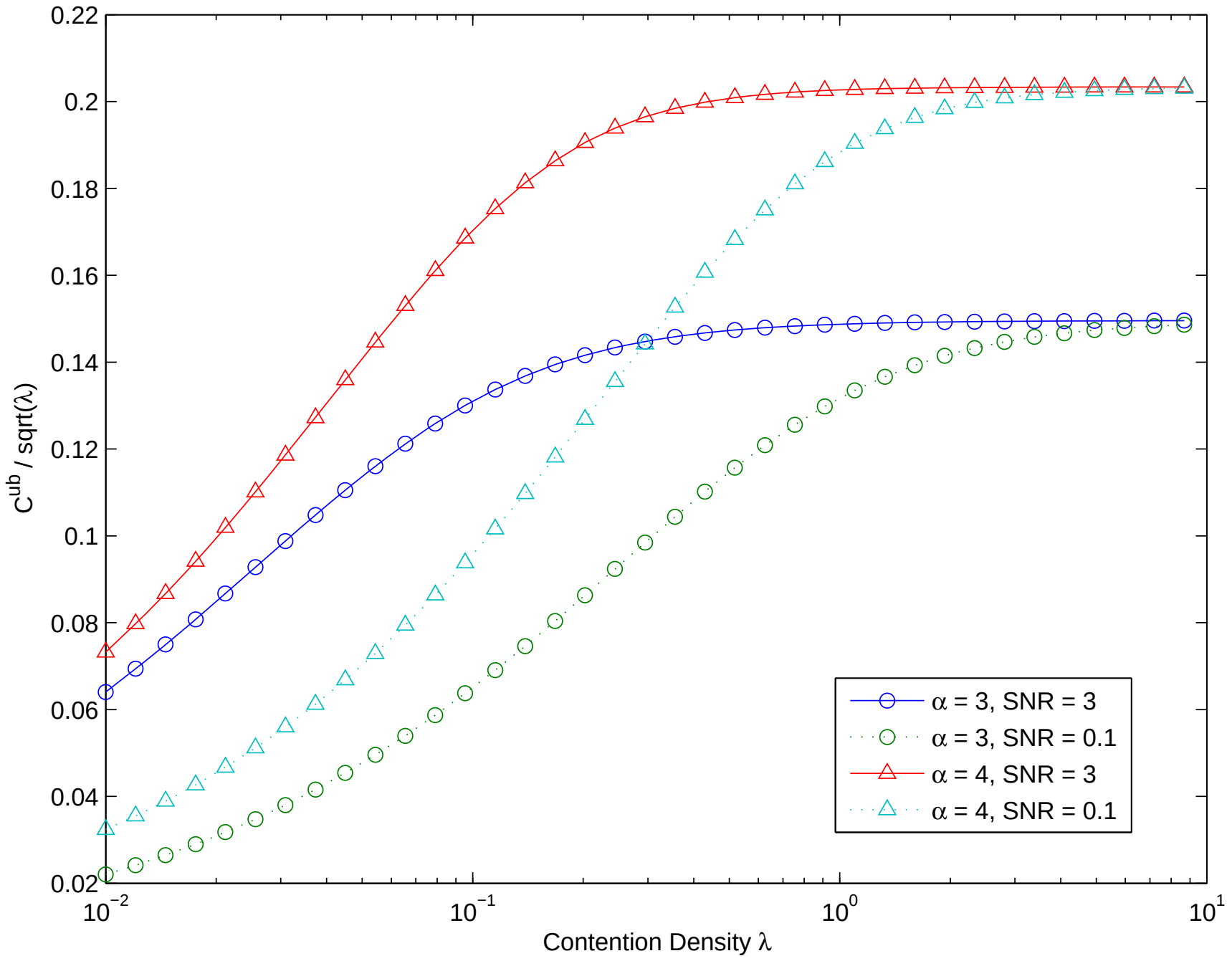

Fig. 7. Normalized random access transport capacity $C^{\mathrm{ub}}/\sqrt{\lambda}$ vs. contention density $\lambda$ when using $M^{*}(\lambda)$ hops. The $\sqrt{\lambda}$ scaling law kicks in more quickly at high SNR.